\documentclass[doublecol]{epl2}

\newcommand{\bn}{\begin{enumerate}}
\newcommand{\en}{\end{enumerate}}
\newcommand{\ba}{\begin{eqnarray}}
\newcommand{\ea}{\end{eqnarray}}

\usepackage{graphicx,color}

\newcommand{\be}{\begin{equation}}
\newcommand{\ee}{\end{equation}}

\newcommand{\et}{{\it et al. }}
\newcommand{\ete}{{\it et al.}}

\def\prl{{ Phys. Rev. Lett. }}

\def\prb{{ Phys. Rev. B }}

\title{A new and simple model for magneto-optics uncovers an
  unexpected spin switching}
\shorttitle{Spin switching} 

\author{G. P. Zhang\inst{1} \and Y. H. Bai\inst{2} \and Thomas F. George\inst{3}}
\shortauthor{G. P. Zhang \etal}

\institute{
  \inst{1}Department of Physics, Indiana State University,
   Terre Haute, IN 47809, USA\\
  \inst{2}Office of Information Technology, Indiana State
  University, Terre Haute, IN 47809, USA \\
\inst{3}Office of the Chancellor and Center for Nanoscience
  \\Departments of Chemistry \& Biochemistry and Physics \& Astronomy
  \\University of Missouri-St. Louis, St.  Louis, MO 63121, USA

}


\abstract{In magneto-optics the spin angular momentum $S_z$ of a sample is
  indirectly probed by the rotation angle and ellipticity, which are
  mainly determined by the off-diagonal susceptibility
  $\chi^{(1)}_{xy}$. A direct and analytic relation between $S_z$ and
  $\chi^{(1)}_{xy}$ is necessary and of paramount importance to the
  success of magneto-optics, but is often difficult to acquire since
  quantum mechanically the relation is hidden in the sum-over-states.
  Here we propose a new and simple model to establish such a much
  needed relation. Our model is based on the Hookean model, but
  includes spin-orbit coupling. Under cw excitation, we show that
  $\chi^{(1)}_{xy}(\omega)$ is indeed directly proportional to $S_z$
  for a fixed photon frequency $\omega$.  Such an elegant relation is
  encouraging, and we wonder whether our model can describe spin
  dynamics as well. By allowing the spin to change dynamically, to our
  surprise, our model predicts that an ultrafast laser pulse can
  induce a spin precession; with appropriate parameters, the laser can
  even reverse spin from one direction to another. This works for both
  the circularly and linearly polarized light. The spin reversal
  window is narrow. These unexpected results closely resemble
  all-optical helicity-dependent magnetic switching found in much more
  complicated ferrimagnetic rare earth compounds.  Therefore, we
  believe that our spin-orbit coupled model may find some important
  applications in spin switching processes, a hot topic in
femtomagnetism.
}

\begin{document}

\maketitle

\section{Introduction}

Magneto-optical Faraday and Kerr techniques are indispensable to
modern magnetism investigations\cite{kryder,meier}. Such a technique
is an ideal tool to investigate both static and dynamic evolutions of
spin\cite{eric,guidoni,umc13,prl00}, since it is a photon-in and photon-out
technique and leaves the sample intact before and after the
experimental measurement.  The origin of the magneto-optics is rooted
in the spin-orbit coupling and exchange
interaction\cite{arg,es}, where the rotational angle and ellipticity
carry the information of the spin moment change.  The classical
understanding is developed through the harmonic oscillator model
(Hookean model), augmented with the Lorentz force term \cite{arg,es},
\be -\frac{d^2 {\bf r}}{dt^2}+2\gamma {\bf v} + \Omega^2 {\bf r} =
\frac{q {\bf E}(t)}{m}+q {\bf B}\times {\bf v}
\label{eq1}
\ee
where ${\bf r}$ and ${\bf v}$
 are the position and velocity of the electron, $\gamma$ is the damping,
 $\Omega$ is the resonance frequency of the harmonic oscillator, $q$
 is the electron charge, ${\bf B}$ is the external magnetic
 field, and  ${\bf E}(t)$ is the electric field of light. The second
 term on the right-hand side is the Lorentz force term and is
the only term that is responsible for
 the magneto-optical effect.

Recently, Hinschberger and Hervieux\cite{hh} extended the harmonic
oscillator model by introducing more higher order terms, so the
nonlinear optical response can be computed as well. The spin-orbit
interaction is included through the current operator, which becomes
another source term in the Maxwell equation. The spin contribution is
lumped into a single parameter that represents the modification due to
the spin-orbit interaction.  Very recently, the same model was used to
develop a quantum theory for the inverse Faraday
effect\cite{battiato2014}.

Despite the great success of the harmonic oscillator model to describe
the magneto-optics, the link between the magnetic spin moment and the
magneto-optical signal can not be easily established from
Eq. (\ref{eq1}) since the harmonic oscillator model contains no
spin. Argyres \cite{arg} provided a hand-waving argument from the
summation over the magnetic electrons by recognizing \ba \int_v d{\bf
  k}&=&(2\pi)^3 \times {\rm no.~ of~ magnetic~ electrons~ per~ volume}\nonumber\\
&=&(2\pi)^3 \frac{M_s}{\frac{1}{2}g\mu_B},  \ea where ${\bf k}$ is the
crystal momentum, $M_s$ is the saturation magnetization, $g$ is the
Land\'e $g$-factor, and $\mu_B$ is the Bohr magneton. Even in modern
magneto-optics band-structure theory, such a relation between the spin
moment and the off-diagonal susceptibility is also hidden in the
summation over the crystal momentum \cite{np09}. An alternative is to
adopt an relaxation-time approximation\cite{es}. Given the importance
of the time-resolved magneto-optical response in laser-induced
ultrafast spin dynamics\cite{eric}, this is highly unsatisfactory in
comparison with linear optics, where the susceptibility can be
directly connected to the dipole transition\cite{blo,boyd}.

In this paper, we make a moderate attempt to establish an analytic
link between the spin angular momentum and off-diagonal susceptibility
in the classical Hookean model. Our treatment is simple and
straightforward, and physically more intuitive.  We first replace the
above Lorentz term by a spin-orbit coupling term. Then, we use the
equation of motion method to find the evolution of the electron
position. Under the continuous wave (cw) excitation, we obtain an analytic
expression for the off-diagonal susceptibility $\chi^{(1)}_{xy}$. We
find that $\chi^{(1)}_{xy}$ is indeed directly proportional to the
product of the spin $S_z$ and the spin-orbit coupling
$\lambda$. Different from the regular linear optical response, the
susceptibility has four poles, instead of two. These additional poles
result from the spin-orbit coupling. Encouraged by this elegant
result, we are curious whether such a simple model could accommodate
spin precession. To our surprise, by allowing spin momentum to change
dynamically with time, we find that this spin-orbit coupled Hookean
model is capable of describing laser-induced ultrafast spin precession
as well. For most laser parameters, our model predicts a spin
canting, but in a narrow parameter range, we find spin switching. This
largely unexpected spin switching has the same character as observed
in bulk ferrimagnets, where both linearly and circularly polarized
light can switch spins, and both the laser duration and field amplitude
have a strong effect on the spin switching. This model might have some
important applications for all-optical helicity-dependent magnetic
switching\cite{stanciu,mangin,lambert}.

The rest of the paper is arranged as follows. Section II is devoted to the
theoretical formalism, where the analytic results for the diagonal and
off-diagonal susceptibilities and the equations of motion for
position, momentum, spin and orbital angular momenta are shown. In
Sec. III, we report the spin switching in our spin-orbit
coupled Hookean model. The conclusion is provided in Sec. IV.

\section{Theoretical formalism}

Interest in time-resolved magneto-optics is fueled by the discovery of
the laser-induced ultrafast spin dynamics in
ferromagnets\cite{eric,prl00}.  The field has grown enormously over
two decades\cite{rasingreview,umc13}, with the discovery of
all-optical helicity-dependent magnetic
switching\cite{stanciu,ostler,mangin,lambert}. Our initial motivation
for this study came from a long discussion in 1999 with one of our
experimental colleagues\cite{bigot} who suggested whether a similar
harmonic oscillator model as in traditional nonlinear optics
\cite{blo} could be developed. However, it was unclear then whether a
model, even if developed, could contain enough physics to be useful to
experimentalists. New investigations have now demonstrated the
potential value in this direction \cite{hh,battiato2014}. One big
issue, however, is how to include the spin-orbit coupling (SOC)
quantum mechanically. In 2011, when we investigated ultrafast spin
linear reversal\cite{jpcm11}, we found a possibility to include SOC
within the Heisenberg model, but when we tried to include the laser
field, a major difficulty appeared: How to couple the laser field to
the system in the Heisenberg model? In particular, the vector
potential of the laser field is spin-independent. In 2014, while
discussing with our colleagues from University of
Colorado\cite{patrik}, we realized that the coupling could also be
treated through the harmonic oscillator model. The only thing left to
be established is the connection among momentum, spin and orbital
angular momenta.  In 1998, to explain how SOC rotates the polarization
plane of linearly polarized light, H\"ubner \cite{wolfgang1} noticed
that an additional force from SOC acts upon the electron through
$i\hbar \dot{\bf p} =[H_{soc}, {\bf p}] =\rm force$. We wondered
whether a similar equation can be derived for the spin and orbital as
well.

However, while testing this idea, it became immediately clear to us
that we ought to replace the Lorentz term in the original harmonic
oscillator model by SOC.  Thus  our Hamiltonian takes the form
\cite{mingsu10,jpcm11,jpcm13,jpcm14,jpcm15} \be H=\frac{{\bf
    p}^2}{2m}+\frac{1}{2}m\Omega^2 {\bf r}^2 +\lambda {\bf L}\cdot
     {\bf S} -e {\bf E}(t) \cdot {\bf r}, \label{ham} \ee where the
     first term is the kinetic energy operator of the electron, the
     second term is the harmonic potential energy operator with system
     frequency $\Omega$, $\lambda$ is the spin-orbit coupling in units
     of eV/$\hbar^2$, $ {\bf L}$ and $ {\bf S} $ are the orbital and
     spin angular momenta in units of $\hbar$, respectively, and {\bf p}
     and {\bf r} are the momentum and position operators of the
     electron, respectively.

To start with, we treat all the operators quantum mechanically and use
the Heisenberg equation of motion \cite{wolfgang1} to find four
coupled equations for the position ${\bf r}$, momentum ${\bf p}$, spin
${\bf S}$ and orbital ${\bf L}$ angular momenta, respectively: \ba
\frac{d\bf r}{dt}&=&\frac{\bf p}{m}-\lambda ({\bf r}\times {\bf
  S})\label{position},\\ \frac{d \bf p}{dt}&=&-m\Omega^2 {\bf r}+e{\bf
  E}(t) -\lambda {\bf p}\times {\bf S} \label{momentum},\\ \frac{d\bf
  S}{dt}&=&\lambda ({\bf L} \times {\bf S})\label{spin},\\ \frac{d\bf
  L}{dt}&=&-e{\bf E}(t) \times {\bf r} - \lambda ({\bf L} \times {\bf
  S}). \label{orbital}\ea Up to this step, these equations are exact. Next we make the
Hartree-Fock approximation by replacing the operators with their
respective expectation values; for simplicity we keep the same
symbols. This allows us to solve these coupled equations analytically.

 These four coupled equations contain incredibly rich physics.  First
 we note that the position's change with time is renormalized by the
 spin-orbit coupling (see Eq. (\ref{position})). This is very
 different from the conventional equation of motion for the electron
 under an external magnetic field {\bf B}, where {\bf B} enters the
 equation through the momentum alone (see Eq. (\ref{eq1})). The
 momentum change also contains a contribution from SOC.  Second, the
 spin does not depend on the laser field directly, but instead on the
 orbital momentum.  {Upon the laser excitation, the orbital
   angular momentum {\bf L} is first excited (Eq. (\ref{orbital})). The
   spin dynamics is driven by the the spin-orbit coupling through a
   spin-orbit torque (Eq. (\ref{spin})). The process is similar to a
   recent study in a magnetic semiconductor by Lingos and coworkers
   \cite{lingos} and in a more complicated magnetic system \cite{li}.
   The module of spin is conserved, or $\dot{\bf S}\cdot {\bf S}=0$,
   as seen from Eq. (\ref{spin}).}

Since we are interested in an analytic solution, we consider a case
with constant spin. We take the derivative of Eq. (\ref{position})
with respect to time and then substitute Eq. (\ref{momentum}) into
it. After some rearrangement, the equation of motion for the electron
position can be simplified to a single equation, \be \ddot{\bf
  r}+2\lambda \dot{\bf r}\times {\bf S}+ (\Omega^2-\lambda^2S^2){\bf
  r}-\lambda^2({\bf r}\cdot{\bf S}){\bf S}=\frac{e{\bf
    E}(t)}{m}.\label{new} \ee This resultant equation closely
resembles the classical Hookean harmonic oscillator equation
(Eq. (\ref{eq1})), but with some differences. The second term on the
left-hand side is not a damping term; instead, it represents a
rotation along an orthogonal direction.  The harmonic potential term
(third term on the left-hand side) is modified with a frequency
red-shifted to $(\Omega^2-\lambda^2 S^2)$, where $S$ is the module of
${\bf S}$.  We align our spin along the $z$ axis or ${\bf
  S}=S_z\hat{z}$.  We apply a cw optical field polarized along the $x$
direction, $E_x(t)=A_x e^{i\omega t}+cc.$, where $A_x$ is the
amplitude and $\omega$ is the laser frequency. From Eq. (\ref{new}),
we find two coupled equations for $x$ and $y$ as \ba \ddot{x}+2\lambda
S_z\dot{y}+(\Omega^2-\lambda^2S_z^2)x&=&\frac{eE_x(t)}{m}\\ \ddot{y}-2\lambda
S_z \dot{x}+ (\Omega^2-\lambda^2S_z^2)y&=&0. \ea To find a stationary
solution, we assume $x(t)=x_0 {\rm e}^{i\omega t}$ and $y(t)=y_0 {\rm
  e}^{i\omega t}$. We substitute them back into Eqs. (7) and (8) to
find \ba x_0= \frac{e}{m}\frac{D}{D^2-C^2}A_x\\ y_0=
\frac{ie}{m}\frac{C}{D^2-C^2}A_x, \ea where
$D=\Omega^2-\omega^2-\lambda^2S_z^2$ and $C=2\lambda S_z\omega$.

To compute the susceptibility, we need to find the polarizations along
the $x$ and $y$ directions \cite{boyd}: $P_x(\omega)=-Nex_0$ and
$P_y(\omega)=-Ney_0$. The diagonal susceptibility (in SI units) is
just $\chi_{xx}^{(1)}=P_x/A_x\epsilon_0$, and the off-diagonal
susceptibility (in SI units) is $\chi_{xy}^{(1)}=P_y/A_x\epsilon_0$,
i.e., \ba \chi_{xx}^{(1)}(\omega)&=&-\frac{Ne^2}{\epsilon_0
  m}\frac{\Omega^2-\omega^2-\lambda^2S_z^2 }{
  (\Omega^2-\omega^2-\lambda^2S_z^2)^2-(2\lambda
  S_z\omega)^2}\label{chi1}
\\ \chi_{xy}^{(1)}(\omega)&=&-i\frac{Ne^2}{\epsilon_0 m}\frac{2\lambda
  S_z\omega}{(\Omega^2-\omega^2-\lambda^2S_z^2)^2-(2\lambda
  S_z\omega)^2}, \label{xy}\ea where $N$ is the number density and
$\epsilon_0$ is the permittivity in the vacuum.  These two equations
are insightful.  Equation (\ref{chi1}) shows that SOC creates two
additional poles on the frequency axis at $\omega=\pm(\Omega-\lambda
S_z)$ and $\omega=\pm(\Omega+\lambda S_z)$.  Without SOC, we recover
the well-known linear susceptibility results \cite{boyd}. Second, the
off-diagonal linear susceptibility $\chi_{xy}^{(1)}$ is indeed
proportional to the spin angular momentum and  spin-orbit
coupling. To the best of our knowledge, such an analytic expression
has never been found before.  It is truly gratifying that our simple
model can reproduce such an elegant relation and sets up a classical
analogue to the linear optics counterpart \cite{boyd,np09}, placing
the entire magneto-optical theory on a firmer ground for the first
time \cite{np09}.

\section{Surprising spin switching}

Being able to analytically connect the off-diagonal susceptibility to
the spin angular momentum is encouraging, but the spin angular
momentum is assumed to be static, though we have four coupled
equations. We are curious as to what happens if we allow the spin to
dynamically change with time under a laser pulse.

We choose laser pulses of two different kinds.  For a linearly
polarized ($\pi$) pulse, the electric field is $ {\bf E}(t)=A_0 {\rm
  e}^{-t^2/\tau^2}\cos(\omega t) \hat{x},$ where $\omega$ is the laser
carrier frequency, $\tau$ is the laser pulse duration, $A_0$ is the
laser field amplitude, $t$ is time, and $\hat{x}$ is the unit vector
along the $x$ axis. Note that the results are the same if the field is
along the $y$ axis. The electric field for the right and left
circularly polarized pulses ($\sigma^{+}$ and $\sigma^{-}$) is $ {\bf
  E}(t)= A_0 {\rm e}^{-t^2/\tau^2} (\pm \sin(\omega t)
\hat{x}+\cos(\omega t) \hat{y})$, where $+(-)$ refers to
$\sigma^{+}$($\sigma^{-}$).  We then compute the spin evolution by
numerically solving the four coupled equations
(\ref{position}-\ref{orbital}).

\begin{figure}
\includegraphics[angle=270,width=8.5cm]{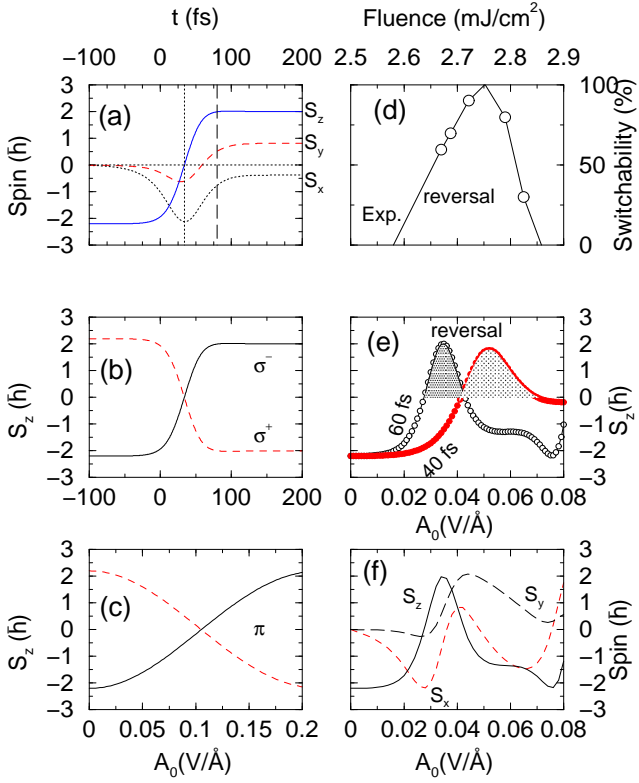}
\caption{ (a) All-optical spin reversal for $S_x$, $S_y$ and $S_z$ as
  a function of time $t$.  The vertical dotted line denotes the time
  when $S_z$ passes through zero, and the vertical dashed line denotes
  the time when the spin reversal starts. Here $\sigma^{-}$ is used,
  the laser pulse duration is $\tau=60$ fs, and the field amplitude is
  0.035 $\rm V/\AA$.  (b) The $\sigma^{-}$ pulse (solid line) only
  switches spin from down to up, while the $\sigma^{+}$ pulse (dashed
  line) only switches spin from up to down. (c) The $\pi$ pulse can
  switch spin from up to down or from down to up, but at a much higher
  field amplitude.  (d) Experimental spin reversal window from Ref. 28.
   (e) Final spin angular momentum $S_z$ as a
  function of the laser field amplitude for laser durations $\tau=60$
  fs (empty circles) and 40 fs (filled circles). The shaded regions
  are the spin reversal window.  (f) As the field amplitude increases,
  the spin angular momentum changes from non-switching, canting along
  the $-x$ axis, switching, and canting along the $+y$ axis.
}
\label{fig1}
\end{figure}

We find that for most sets of laser parameters the spin precesses with
time, but for a few special set of laser parameters, the laser can
directly switch spin, without the presence of a magnetic field. This
is a big surprise to us.  Figure \ref{fig1}(a) shows that a
left-circularly polarized pulse ($\sigma^{-}$) of duration $\tau=60$
fs and field amplitude $A_0=0.035\rm V/\AA$ can switch the spin,
initialized along the $-z$ axis with $S_z(0)=-2.2\hbar$, to $2\hbar$.
{We choose the spin-orbit coupling $\lambda=0.06{\rm
    eV}/\hbar^2$, and $\hbar\omega=\hbar\Omega=1.6$ eV.  Note that our
  excitation is slightly off-resonant since new poles come from SOC
  (see Eq. (\ref{xy})).}  We find that upon the laser excitation, the
spin first precesses strongly without any oscillation toward the $xy$
plane and the exact precession of $S_x$ and $S_y$ depends on the
initial phase of the laser pulse, but the precession of $S_z$ is
always the same.  The spin reaches the negative maximum at 34 fs,
exactly when $S_z$ passes through zero. $S_z$ is successfully switches
to $2\hbar$ at 80 fs, where the spin rotates 155.9$^\circ$.  Such a
spin switching is remarkable, and to the best of our knowledge, has
never been reported before.  Will this work for a $\sigma^+$ pulse?

 Figure \ref{fig1}(b) shows that both $\sigma^{+}$ and $\sigma^{-}$
 can switch spin within a few hundred femtoseconds.  Our results
 reveal a stringent symmetry constraint on the spin switching: The
 $\sigma^{-}$ light only switches the spin from down to up, while the
 $\sigma^{+}$ light switches the spin from up to down, not the other
 way around.  Numerically we find that $\sigma^{+}$ only slightly
 perturbs the down spin, because the phase mismatch between the
 position of the electron and the laser field imposes a negative
 torque on the spin switching. We also try to use linearly polarized
 light ($\pi$).  Figure \ref{fig1}(c) shows that for the same
 amplitude of 0.035$\rm V/\AA$, the $\pi$ pulse can not switch the
 spin, and only a small spin change is noticed.  To induce spin
 switching, we need to increase $A_0$ above 0.2$\rm V/\AA$, or 5.7
 times higher than used for either $\sigma^{+}$ or $\sigma^{-}$.

Our finding reminds us some of familiar experimental results found in
all-optical helicity-dependent magnetic switching in ferrimagnetic
rare-earth bulk materials \cite{mangin,lambert}. For instance,
experimentally, Stanciu \et \cite{stanciu} showed that both
$\sigma^{+}$ or $\sigma^{-}$ can switch spins; and Alebrand \et
\cite{sabine2012} found that a $\sigma$ pulse appears more powerful
than a $\pi$ pulse. In addition, Vahaplar \et \cite{vahaplar} found
that the actual spin reversal window of the laser fluence is very
narrow and asymmetric (see Fig. \ref{fig1}(d)). We wonder whether a
similar switch window exists for our system.  Figure \ref{fig1}(e)
shows that as $A_0$ increases, the final spin $S_z$ first increases
sharply (see the empty circles) and then reaches its maximum of
$2\hbar$ at $A_0=0.035$$\rm V/\AA$, where the spin is reversed.  If we
increase the field amplitude further, $S_z$ decreases and eventually
the spin switching disappears.  The reversal window is indeed very
narrow and asymmetric (see the shaded region in Fig. \ref{fig1}(b)),
only from 0.026 to 0.042 $\rm V/\AA$.

To understand how such a narrow reversal window is formed, we
systematically monitor the orbital and spin angular momentum changes
as a function of time. We find that while the orbital angular momentum
always increases with the field amplitude, the spin change is
nontrivial.  If the laser is too weak, weaker than 0.026 $\rm V/\AA$,
the spin either does not rotate at all or simply cants to the $-x$
axis, without switching. If $A_0$ is too strong, between 0.035 $\rm
V/\AA$ and 0.06 $\rm V/\AA$, the final spin overshoots and cants to
the $+y$ axis. It is this balance between these two limits that leads
to the narrow reversal window. Figure \ref{fig1}(f) shows the detailed
dependence of the final spin components as a function of
amplitude. Note that the actual dependence at the higher amplitude is
very complicated, and further spin switching and canting are possible.
If we shorten the pulse duration, we find that the switching window is
shifted to the high intensity end and is slightly widened (see
Fig. \ref{fig1}(e)).  This shift is expected since to switch the spin,
the laser field must transfer enough energy and angular momentum to
the system.  If the pulse is too short, the spin does not have enough
time to reverse before the laser pulse is gone.

\begin{figure}
\includegraphics[angle=270,width=8.8cm]{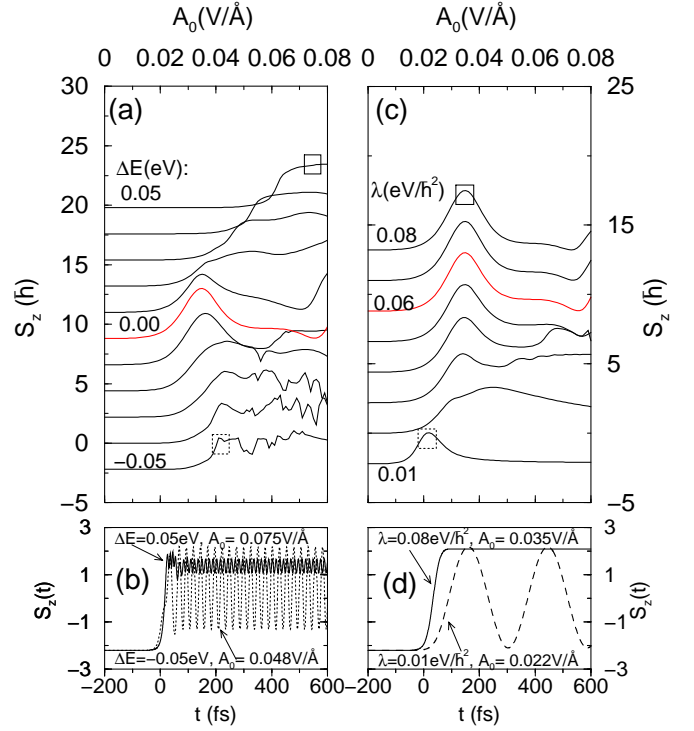}
\caption{{ (a) Dependence of the spin reversal window on the
    energy difference $\Delta E$ with fixed $\lambda=0.06 {\rm
      eV}/\hbar^2$, where $\Delta E=\hbar \omega -\hbar\Omega$.
    $\Delta E$ increases from -0.05 to 0.05 eV in steps of 0.01 eV
    from the bottom to top. All the curves are shifted vertically for
    clarity. The solid and dashed boxes denote two representative
    cases, whose spin change is shown in (b).  (b) Spin angular
    momentum change with time for the laser field amplitude of
    $A_0=0.075\rm V/\AA$ and $\Delta E=0.05$ eV (solid line) and
    $A_0=0.049\rm V/\AA$ and $\Delta E=-0.05$ eV (dashed line).  (c)
    Dependence of the spin reversal window on the spin-orbit coupling
    $\lambda$, with $\Delta E$ fixed at 0 eV. Here $\lambda$ increases
    from 0.01 to 0.08 eV/$\hbar^2$ in steps of 0.01 eV/$\hbar^2$.
    (d) Spin angular momentum change with time for
 $A_0=0.035\rm V/\AA$ and $\lambda=0.08{\rm eV}/\hbar^2$
    (solid line) and $A_0=0.022\rm V/\AA$ and $\lambda=0.01{\rm
      eV}/\hbar^2$ (dashed line).
}}
\label{fig2}
\end{figure}

{ Up to now, all the results are obtained with a fixed
  $\lambda$ and $\hbar \omega$. Next we tune them separately. We first
  fix $\lambda$ at 0.06 eV$/\hbar^2$.  Figure \ref{fig2}(a) shows that
  the energy difference ($\Delta E=\hbar\Omega-\hbar\omega$) between
  the laser photon energy and the system energy affects the spin
  reversal window strongly. We increase $\Delta E$ from -0.05 to 0.05
  eV in steps of 0.01 eV.  We see that at $\Delta E=-0.05$ eV, the
  spin reversal window is broader but there is no spin reversal. To
  see this clearly, we choose a point with an optimal field amplitude
  (see the small dashed box in Fig \ref{fig2}(a)), and plot the spin
  angular momentum as a function of time in Fig \ref{fig2}(b) (see the
  dashed line).  We notice that the spin only oscillates between two
  extremes (see the dashed line in Fig \ref{fig2}(b)), and does not
  reverse even at this optimal laser amplitude.  However, as we
  increase $\Delta E$ further close to 0 eV, the spin reversal window
  sharpens, with the peaks on the high end of the laser field
  amplitude dropping off, and spin reversal starts.  Our result at
  $\Delta E=0$ eV is highlighted in red.  Above $\Delta E=0$ eV, the
  reversal window broadens and forms a plateau. This means that the
  spin reversal can occur at different $A_0$.  We also choose a point
  at $\Delta E=0.05$ eV (see the solid box), and plot the spin change
  in Fig. \ref{fig2}(b) (solid line), where we see that the spin is
  indeed reversed, but with a small ringing.

Next we fix $\Delta E$ at 0 eV, and increase $\lambda$ from 0.01 to
$\lambda=0.08~{\rm eV}/\hbar^2$, in steps of $0.01~{\rm eV}/\hbar^2$.
Figure \ref{fig2}(c) shows that at $\lambda=0.01~{\rm eV}/\hbar^2$,
even with an optimal amplitude (see the dashed box in
Fig. \ref{fig2}(c)), the spin only oscillates and does not reverse
(see the dashed line in Fig. \ref{fig2}(d)).  As $\lambda$ increases,
the spin reversal window also changes its shape, and the optimal field
amplitude gradually shifts to a high value and then converges.  A
minimum $\lambda_{min}$ is required to reverse the spin.  We estimate
that the minimum $\lambda_{min}$ to reverse spin is around
$\lambda=0.03~{\rm eV}/\hbar^2$ for our current parameters. Figure
\ref{fig2}(d) also shows a successful spin reversal for our largest
$\lambda=0.08~{\rm eV}/\hbar^2$ (see the solid line). The entire
process is similar to our results at $\lambda=0.06~{\rm eV}/\hbar^2$.
At present, we are still investigating the possibility to develop a
simple physics picture from Eqs. (4)-(7). }

So many similarities in the spin reversal between our spin-orbit
coupled harmonic oscillator and ferrimagnetic rare earth bulks are
astonishing and highly unexpected. We wonder whether they share the
common origin. Physically, they both have the spin-orbit coupling, are
coupled with the laser field, and do not rely on an external magnetic
field. One apparent difference between them is that in our system the
spin reversal is much faster with the presence of the laser pulse, but
in ferrimagnets on the order of several picoseconds when the laser
field is gone. However, this difference may be explained by the size
effect. Our system contains one single site; regular ferrimagnets have
a lot more. It could be that in those ferrimagnets, the spin reversal
initially occurs in the small ranges with the strongest laser field
amplitude, and the propagation of the spin reversal, from those initial
excited sites to the remote sites, takes several picoseconds
\cite{ostler}. {Nevertheless, we caution that these
  ferrimagnets consist of two sublattices and are quite different from
  our model.}  Additional research is needed in this direction to
completely illuminate the true mechanism.

\section{Conclusion}

We have proposed a new and simple spin-orbit coupled harmonic
oscillator model for magneto-optics to establish the analytic relation
between the experimentally accessible off-diagonal susceptibility
$\chi^{(1)}_{xy}(\omega)$ and experimentally inaccessible spin angular
momentum $S_z$. We show that under cw excitation, for the same photon
frequency $\omega$, $\chi^{(1)}_{xy}(\omega)$ is directly proportional
to the product of the spin-orbit coupling ($\lambda$) and the spin
moment $(S_z)$, or $\lambda S_z$. Different from the traditional
optical response, the off-diagonal susceptibility has four poles on
the frequency axis, instead of two. These two extra poles are a
result of spin-orbit coupling. Therefore, our results provide an
important theoretical foundation for magneto-optics, in contrast to
the prior efforts, where such a relation is obtained approximately or
hidden in the sum-over-states. To our surprise, our model, once the
spin is allowed to change and under the laser excitation, predicts the
laser-induced ultrafast spin precession; with appropriate laser
parameters, the spin reverses from one direction to another. This works for both the
circularly and linearly polarized light, but the former appears
more powerful than the latter. The spin reversal window is very narrow
due to the stringent requirement on the precession of the spin. All
these results are similar to the all-optical helicity-dependent
magnetic switching (AOS) observed in ferrimagnets. At present, it is
unknown whether the same physics plays the role in both systems. We
believe that an extension of our current model to a large system could
be fruitful to AOS, where the research has been very intensive, with
the possible applications in ultrafast magnetic storage as
demonstrated in several latest investigations \cite{mangin,lambert}.

\acknowledgments This work was solely supported by the U.S. Department
of Energy under Contract No. DE-FG02-06ER46304. Part of the work was
done on Indiana State University's quantum cluster and
high-performance computers.  The research used resources of the
National Energy Research Scientific Computing Center, which is
supported by the Office of Science of the U.S. Department of Energy
under Contract No. DE-AC02-05CH11231. This work was performed, in
part, at the Center for Integrated Nanotechnologies, an Office of
Science User Facility operated for the U.S. Department of Energy (DOE)
Office of Science by Los Alamos National Laboratory (Contract
DE-AC52-06NA25396) and Sandia National Laboratories (Contract
DE-AC04-94AL85000).

\end{document}